\documentclass[prl,aps,amssymb,amsmath,
superscriptaddress,twocolumn,showpacs]{revtex4}

\renewcommand{\epsilon}{\varepsilon}

\newcommand{\cE}{\mathcal{E}}
\newcommand{\cT}{\mathcal{T}}

\renewcommand{\phi}{\varphi}
\newcommand{\R}{\mathbb{R}}

\newcommand{\br}{\mathbf{r}}
\newcommand{\bp}{\mathbf{p}}
\newcommand{\bq}{\mathbf{q}}
\newcommand{\bk}{\mathbf{k}}

\DeclareMathOperator{\Tr}{Tr}
\DeclareMathOperator{\tr}{Tr}

\begin{document}

\title{Energy Cost to Make a Hole in the Fermi Sea}

\author{Rupert L. Frank}
\affiliation{Department of Mathematics,
Princeton University, Washington Road, Princeton, NJ 08544, USA}

\author{Mathieu Lewin}
\affiliation{CNRS and Department of Mathematics (UMR CNRS 8088), 
University of Cergy-Pontoise, F-95000 Cergy-Pontoise, France}

\author{Elliott H. Lieb}
\affiliation{Department of Mathematics,
Princeton University, Washington Road, Princeton, NJ 08544, USA}
\affiliation{Department of Physics,
Princeton University, P.~O.~Box 708, Princeton, NJ 08542, USA}

\author{Robert Seiringer}
\affiliation{Department of Mathematics and Statistics, McGill University, 
805 Sherbrooke Street West, 
Montr\'eal, Qu\'ebec, H3A 2K6,
Canada }

\begin{abstract}
  The change in energy of an ideal Fermi gas when a local one-body
  potential is inserted into the system, or when the density is
  changed locally, are  important quantities in condensed
  matter physics.  We show that they can be rigorously
  bounded from below by a universal constant times the value given by the
  semiclassical approximation.\\
Final version published in \textit{Phys. Rev. Lett.} \textbf{106}, 150402 (2011).
\end{abstract}


\maketitle

A problem of long-standing interest in condensed matter physics is to
give an effective estimate of the minimum change in kinetic energy,
$\delta\cT(\delta\rho)$, of an infinitely extended ideal Fermi gas
when the density is changed locally by a fixed, specified amount
$\delta\rho(\br)$, i.e., the density is changed from a constant
$\rho_0$ to $\rho(\br)=\rho_0+\delta\rho(\br)$.  Note that
$\delta\rho(\br)$ can be negative, as long as $\delta\rho(\br)\geq
-\rho_0$, hence the word ``hole'' in our title.

An equivalent problem is to calculate the minimum change in total
energy $\delta\cE(V)$ of the Fermi gas when a local, one body
potential $V(\br)$, of either sign, is added to the kinetic energy
Hamiltonian, i.e., $-\nabla^2 \to -\nabla^2 +V(\br)$, where we use
units such that $\hbar = 2m = 1$.  This rigorous equivalence, recalled
in \eqref{eq:Legendre-rho} and \eqref{eq:Legendre-V} below, is well known in
density-functional theory.

In this Letter we will give an effective answer to both questions by
proving rigorously for dimensions $D\geq 2$ that the well known
semiclassical approximations are, up to an overall constant, a lower
bound for the kinetic energy cost, as well as for the energy shift
caused by $V(\br)$. In the semiclassical approximation one associates
one quantum state with each box $\Delta \bp \Delta \br$ in phase space
of volume $2\pi$, but this calculation is qualitatively wrong for $D=1$
because of a singularity at the Fermi surface related to the Peierls
instability. 

Plainly, there can not be an upper bound to $\delta\cT(\delta\rho)$
because we can always put the particles in high momentum states while
keeping $\rho(\br)$ fixed. The interesting computational question is
the lower bound, i.e., the minimum required payment for a perturbation
of $\rho_0$.

The literature on $\delta\cE(V)$ is mostly based on perturbation theoretic
ideas. Such calculations are valid in many important cases but they do not
bring the physics to the foreground as sharply as the semiclassical formula
does. It is, therefore, important, conceptually and computationally, to be able
to view the physically transparent semiclassical formula as yielding a
rigorous, non-perturbative bound. The semiclassical
formulas are strictly local in position-space, and hence additive over
impurities, making them useful for density-functional theory \cite{HoKo}.
No multiple-scattering calculation is needed here.

The two energy shifts, $\delta T(\delta \rho)$ and $\delta\cE(V)$, are connected
via a Legendre transform as
\begin{align}\label{eq:Legendre-rho}
  \delta\cT(\delta\rho)&=\sup_{V(\br)}\!\left(\delta\cE(V)-
    \int_{\R^3}\! V(\br)(\rho_0+\delta\rho(\br))\,d^3r\right), \\
  \delta\cE(V)&=\!\!\inf_{\delta\rho(\br)
}\!
  \left(\delta\cT(\delta\rho)+\!\int_{\R^3}\!
    V(\br)(\rho_0+\delta\rho(\br))d^3r\right).
\label{eq:Legendre-V}
\end{align}

If $\rho_0\equiv0$ then $\rho(\br)=\delta\rho(\br)\geq 0$ and
$\cT(\rho)= \delta\cT(\delta\rho)$. In this case we are just creating
a pile of $N$ electrons with density $\rho(\br)$ and with
$\int_{\R^3}\rho(\br)\,d^3r=N$ or, equivalently, we are filling the
negative energy states of a potential $V$. The 3D semiclassical
(also known as Thomas-Fermi) energies are
\begin{align} 
  \cT_{\rm sc}(\rho)&=(3/5)(6\pi^2/q)^{\frac 23}
  \int_{\R^3}\rho(\br)^{\frac 53}\,d^3r,   \label{eq:semi-T}  \\
  \cE_{\rm sc}(V)&=-(q/15\pi^2)\int_{\R^3}V(\br)_-^{\frac 52}\,d^3r,
  \quad\text{if}\ \rho_0\equiv0,
\end{align}
where $y_\pm\equiv \max(0,\pm y)\geq0$ is the positive or negative
part of a number $y$, and $q$ is the number of available spin states per
particle, which is 2 for unpolarized electrons.

When $\rho_0>0$, the semiclassical quantities in 3D are
\begin{align}
  \delta\cT_{\rm sc}(\delta\rho)&=\cT_{\rm
    sc}(\rho_0+\delta\rho)-\cT_{\rm sc}(\rho_0
  )-\mu\;\int_{\R^3}\delta\rho(\br)\,d^3r\nonumber\\
  &=\frac35\left(\frac{6\pi^2}{q}\right)^{\frac 23}
  \int_{\R^3}\bigg\{\left(\rho_0+\delta\rho(\br)\right)^{\frac 53}-\rho_0
  ^{\frac 53}\nonumber\\
  &\quad\qquad\qquad\qquad\qquad-\frac{5}{3}\rho_0^{\frac 23}\delta\rho(\br)\bigg\}\,d^3r,
\label{eq:def_delta_T_sc}  \\
\delta\cE_{\rm sc}(V)&=\cE_{\rm sc}(V-\mu)-\cE_{\rm sc}(-\mu)
\nonumber\\
&=-\frac{q}{15\pi^2} \int_{\R^3}
\bigg\{(V(\br)-\mu)_-^{\frac 52}-\mu^{\frac 52}\bigg\}\,d^3r, 
\end{align}
where $\mu = (6\pi^2\rho_0/q)^{\frac 23}$ is the chemical potential at density $\rho_0$.

\medskip

\noindent\textbf{Main theorem.}
Our main Theorems in 3D are
\begin{equation}
\boxed{\phantom{\Big|} \delta\cT(\delta\rho)
\geq \;0.1279\; \delta\cT_{\rm sc}(\delta\rho)\phantom{\Big|}}
\label{eq:LT}
\end{equation}
for the change in kinetic energy, and
\begin{equation}
\boxed{
\begin{array}{l}
  \displaystyle 0\geq \delta\cE(V) - 
\rho_0\;\int_{\R^3} V(\br)\,d^3r\\[10pt]
  \displaystyle \qquad\quad \geq \;21.85\;
  \left(\delta\cE_{\rm sc}(V)- \rho_0\;\int_{\R^3} V(\br)\,d^3r\right)
\end{array}}
\label{eq:LT-dual}
\end{equation}
for the change in energy when an \textit{arbitrary} local potential
$V(\br)$ is inserted into the system. Similar results hold for all
$D\geq 2$. Our bound~\eqref{eq:LT-dual} quantifies the validity of
first-order perturbation theory, since $\rho_0\;\int_{\R^3}
V(\br)\,d^3r$ is precisely the first-order term. In
\eqref{eq:def_delta_T_sc} note the $(\delta\rho)^{\frac 53}$ dependence for
large $\delta\rho$, as in \eqref{eq:semi-T}, but $(\delta\rho)^2$ for small
$\delta\rho$.

A lower bound in the $\rho_0\equiv0$ case was provided by the
Lieb-Thirring inequality~\cite{LieThi-75,Lieb-76,LieSei-10,DLL} for
\textit{all} $D\geq 1$. For $D=3$
\begin{equation}
  \cT(\rho)= \delta\cT(\delta\rho) 
\geq 0.6724\; \cT_{\rm sc}(\rho), \quad\text{when}\ \rho_0\equiv0.
\label{eq:usual-LT}
\end{equation}
It is widely believed that $0.6724$ can be replaced by $1$, and there is
continuing research in this direction. 
The inequality $\cT(\rho)\geq
K\,\cT_{\rm sc}(\rho)$, as in \eqref{eq:usual-LT},  is equivalent to
\begin{equation}
\cE(V)\geq K^{-3/2} \;\cE_{\rm sc}(V),  
\quad\text{when}\ \rho_0\equiv0.
\label{eq:usual-LT-dual}
\end{equation}

The inequality~\eqref{eq:usual-LT} was derived by first
proving~\eqref{eq:usual-LT-dual} and then using the
equivalence~\eqref{eq:Legendre-rho}--\eqref{eq:Legendre-V}.  Our
attempt to follow this route in the positive density case
was not successful. The
situation changed when Rumin~\cite{Rumin} found a way to prove
directly the kinetic energy bound~\eqref{eq:usual-LT}. By suitably
modifying his method, we are now able to derive a lower bound for
$\delta\cT(\delta\rho)$, and consequently on $\delta\cE(V)$, when $\rho_0 > 0$.

Several papers, e.g.~\cite{Friedel-01,DeWitt-56},
deal with this problem from different points of view. In almost all cases, it
has been approached from the side of computing the energy shift with a
given potential $V$. The idea of computing the shift caused by a given
change in density does not seem to have been widely considered. 

If one fixes the particle number $N$ in a very large box and
calculates
the shift in energy caused by $V$, the answer depends
on the box shape and boundary
conditions
\cite{DeWitt-56,Anderson-10}. We are able to avoid these
problems and go directly to the thermodynamic limit by 
\textit{fixing the chemical potential $\mu$} and working in
infinite space. Then we
can look at the unperturbed Hamiltonian $H_0=-\nabla^2-\mu$ or the perturbed one
$H_V=-\nabla^2-\mu+V(\br)$ and fill all the negative energy states, i.e., all
states below the Fermi level. No box is required in our approach; these
Hamiltonians are defined on the whole space $\R^3$. The individual energies are
necessarily infinite but the difference is finite, as we will explain.

To put this more precisely,
we set $\Pi_V$, respectively $\Pi_0$, equal to the
projections onto the negative spectrum of $H_V$ and $H_0$. The change
in energy is then 
\begin{equation}
\delta\cE(V)=\tr\;(H_V\Pi_V-H_0\Pi_0)
\label{eq:def_delta_E}
\end{equation}
where $\Tr$ denotes the trace.

To define the related kinetic energy shift, we consider a one-particle
density matrix $\gamma(\br,\br')$ (suppressing spin indices for simplicity) 
and compute
\begin{equation}
\delta\cT(\delta\rho)=\inf_\gamma\tr\;H_0(\gamma-\Pi_0).
\label{eq:def_delta_T}
\end{equation}
On the right side, we take the infimum over all density matrices
$\gamma(\br,\br')$ whose diagonal is
$\gamma(\br,\br)=\rho_0+\delta\rho(\br)$. The Fermi statistics enters
via the condition on $\gamma$: It is known~\cite{LieSei-10} that the
necessary and sufficient condition for a one-body $\gamma$ to come
from an $N$-body fermionic density matrix is that $0\leq\gamma\leq1$,
as an operator inequality. The same condition is imposed in
\eqref{eq:def_delta_T}.  Note that $\delta\cT(\delta\rho)$ includes
the chemical potential $\mu$ in its definition (since
$H_0=-\nabla^2-\mu$). With this choice, $\delta\cT(\delta\rho)$ is
always positive, regardless of the sign of $\delta\rho(\br)$.
Formally, the reason for this is that $\Pi_0$ is the minimizer of $\tr
H_0\gamma$ among all one-body density matrices $\gamma$.

\medskip\noindent \textbf{Derivation of the lower
  bound~(\ref{eq:LT}):} To simplify the notation, we shall assume that
$q=1$ from now on. The general case is analogous.  Referring to
Eq.~\eqref{eq:def_delta_T}, we consider a density matrix
$\gamma(\br,\br')$ whose density is $\rho_0+\delta\rho(\br)$. We have
to study $Q=\gamma-\Pi_0$, which we write as
$Q=Q^{++}+Q^{--}+Q^{+-}+Q^{-+}$, where $Q^{--}=\Pi_0 Q \Pi_0$,
$Q^{-+}=\Pi_0 Q (1-\Pi_0)$, etc. In Fourier space this means that
$\widehat{Q^{--}}(\bp,\bq)=\Theta(p^2<\mu) \widehat{Q}(\bp,\bq)
\Theta(q^2<\mu)$, etc., with $\Theta$ the Heaviside step function.
The total change $\delta\rho(\br)$ of the density equals the sum of
the densities of each of these terms, e.g.,
$\rho^{++}(\br)=Q^{++}(\br,\br)$, and so on.  Since $0\leq\gamma\leq1$
in the sense of operators, we have $Q^{++}\geq0$ and $-\Pi_0\leq
Q^{--}\leq0$, hence $\rho^{++}(\br)\geq0$ and $-\rho_0\leq
\rho^{--}(\br)\leq0$. However, $\rho^{+-}(\br)=\rho^{-+}(\br)$ has no
sign \emph{a priori}.

The kinetic energy of the diagonal terms $Q^{\pm\pm}$ can be bounded
using the method of \cite{Rumin}. The starting point is the representation
\begin{align}
\tr(H_0Q^{++})&=\tr(|H_0|\,Q^{++})=\int_0^\infty dE\, 
\tr(Q^{++}_E)\nonumber\\
&=\int_{\R^3}d^3r \int_0^\infty dE\;\rho^{++}_E(\br) \label{eq:Rumin}
\end{align}
where $Q_E^{++}=P_{\geq E} Q^{++} P_{\geq E}$, 
$\rho^{++}_E(\br)=Q^{++}_E(\br,\br)$, and $P_{\geq E}$ is the spectral
projection of $|H_0|=|-\nabla^2-\mu|$ onto energies $\geq
E$. By Schwarz's inequality and  $Q^{++}\leq  1$,
\begin{align*}
  &\sqrt{\langle\psi|Q^{++}|\psi\rangle}\\
  &\qquad\leq \sqrt{\langle\psi|P_{\geq E}Q^{++}P_{\geq E}|\psi\rangle}+\sqrt{\langle\psi|P_{\leq E}Q^{++}P_{\leq E}|\psi\rangle}\\
  &\qquad\leq \sqrt{\langle\psi|P_{\geq E}Q^{++}P_{\geq
      E}|\psi\rangle}+\sqrt{\langle\psi|P_{\leq E}|\psi\rangle}
\end{align*}
for any $\psi$. By taking $\psi$ to be a $\delta$-function we obtain
$\sqrt{\rho^{++}(\br)}\leq \sqrt{\rho^{++}_E(\br)}+\sqrt{r(E)}$, where
$r(E)$ is the (spatial) constant density of $P_{\leq E}$, which is
easily found to be
$$r(E)=\frac{1}{6\pi^2}\left((\mu+E)^{3/2}-(\mu-E)_+^{3/2}\right).$$
When we insert this bound on
$\rho^{++}_E(\br)$ into \eqref{eq:Rumin} we obtain
\begin{equation}
\tr(H_0 Q^{++})\geq \int_{\R^3} F(\rho^{++}(\br))\, d^3r
\label{eq:Rumin-diag} 
\end{equation}  
with 
\begin{equation}
F(y)=\int_0^\infty\,dE \left(\sqrt{|y|}-\sqrt{r(E)}\right)_+^2.
\label{eq:def-F}
\end{equation} 
The function $F(y)$ is convex (because
$y\mapsto (\sqrt{|y|}-C)_+^2$ is convex) and behaves like the
semiclassical counterpart in \eqref{eq:def_delta_T_sc} for small and
large $y$.  The kinetic energy of $Q^{--}$ satisfies the same
inequality as \eqref{eq:Rumin-diag}. Using the convexity of $F$ 
we obtain the bound
\begin{equation}
  \tr(H_0 Q)\geq 2\int_{\R^3} F\left(\frac{\rho^{++}(\br)+
\rho^{--}(\br)}{2}\right)\, d^3r.
\label{eq:Rumin-diag-final} 
\end{equation}

For a different bound we consider the off-diagonal terms
$\rho^{+-}=\rho^{-+}$.  Calculating in momentum space and using
Schwarz's inequality
\begin{align}
  &(2\pi)^{\frac 32} \int\! |\rho^{+-}(\br)|^2\,d^3r =(2\pi)^{\frac 32}\int \!
  \rho^{+-}(\br)Q^{+-}(\br,\br)\,d^3r
  \nonumber \\ 
  &=\int_{p^2\leq\mu}d^3p\int_{q^2\geq\mu}d^3q \;\widehat{\rho^{+-}}(\bp-\bq)\,\widehat{Q}(\bp,\bq)\nonumber \\
  &\leq \left(\int_{p^2\leq\mu}d^3p\int_{q^2\geq\mu}d^3q \;\frac{|\widehat{\rho^{+-}}(\bp-\bq)|^2}{|p^2-\mu|^{\frac12}|q^2-\mu|^{\frac12}}\right)^{\frac 12}\label{eq:estim-off-diag} \\   
  &\times \left(\int_{p^2\leq\mu}\!\!d^3p\int_{q^2\geq\mu}\!\!d^3q
    \;|\widehat{Q}(\bp,\bq)|^2|p^2-\mu|^{\frac12}|q^2-\mu|^{\frac12}\right)^{\frac 12}. \nonumber  
\end{align}
The first square root factor on the right side can be rewritten as
\begin{equation}
  \left( \int_{\R^3}\Phi(k)\;|\widehat{\rho^{+-}}(\bk)|^2\; d^3k
\right)^{\frac12},
\end{equation}
where
\begin{equation}
\Phi(k)=\int_{\substack{p^2\leq\mu\\
|\bp-\bk|^2\geq\mu}}\frac{d^3p}{\sqrt{\mu-p^2}\sqrt{|\bp-\bk|^2-\mu}} \,.
\label{eq:def_Phi}
\end{equation}

Our last task is to bound $\Phi(k)$ from above. As a function of $k$, $\Phi(k)$
can be shown to be monotone decreasing. Thus, it attains its maximum
at $k=0$ where it has the value $\Phi(0)=\pi^2\sqrt{\mu}$. We
deduce from \eqref{eq:estim-off-diag} that
\begin{align}
&\frac{8\pi}{\sqrt\mu}\int_{\R^3} |\rho^{+-}(\br)|^2\,d^3r
\label{eq:off-diag-last} \\
&\leq\int_{p^2\leq\mu}\!\!d^3p\int_{q^2\geq\mu}\!\!d^3q \;|
\widehat{Q}(\bp,\bq)|^2|p^2-\mu|^{1/2}|q^2-\mu|^{1/2}.\nonumber
\end{align}
To understand the right side, we recall that $Q=\gamma-\Pi_0$ where
$0\leq\gamma\leq1$.  Hence
\begin{align*}
Q^2=(\gamma-\Pi_0)^2&=\gamma^2- \gamma\Pi_0-\Pi_0\gamma+\Pi_0\\
&\leq\gamma- \gamma\Pi_0-\Pi_0\gamma+\Pi_0=Q^{++}-Q^{--}
\end{align*}
and, therefore, 
$$\tr H_0 Q\geq \tr|H_0|Q^2\geq 2\times\text{(right side of
 \eqref{eq:off-diag-last})}.$$ 
We deduce that
\begin{equation}
\tr H_0 Q\geq \frac{16\pi}{\sqrt\mu}\int_{\R^3} |\rho^{+-}(\br)|^2\,d^3r.
\label{eq:final-off-diag} 
\end{equation}

So far we have found two lower bounds, (\ref{eq:Rumin-diag-final}) and
(\ref{eq:final-off-diag}), for the shift in kinetic energy
$\delta\cT(\delta\rho)=\inf \tr H_0Q$, which we average with coefficients $t$
and $1-t$. Thus, $\delta\cT(\delta\rho) $ is bounded below
by
\begin{equation}
 \int_{\R^3}  \left[2t
F\left(\tfrac{\rho^{++}(\br)+\rho^{--}(\br)}
2\right) +
\tfrac{16\pi}{\sqrt\mu}(1-t)
|\rho^{+-}(\br)|^2\right]\,d^3r,
\label{eq:final-estimate-max}
\end{equation}
where $F$ is in \eqref{eq:def-F}. We must now give a
lower bound to the right side of
\eqref{eq:final-estimate-max} in terms
of the total change in density
$\delta\rho(\br)=\rho^{++}(\br)+\rho^{--}(\br)+2\rho^{-+}(\br)$. These
three quantities are not known separately but they do satisfy the constraints
that $\rho^{++}(\br)+\rho^{--}(\br)\geq-\rho_0$ and
$\delta\rho(\br)\geq-\rho_0$. We then look for a $0\leq t\leq 1$ such that
\begin{multline*}
2t\,F\left(\frac{x}{2}\right)+\frac{16\pi}{\sqrt\mu}(1-t)y^2\\
\geq \kappa\,\frac35\left(\frac{6\pi^2}{q}\right)^{\frac23} \bigg\{\left(\rho_0+x+2y\right)^{\frac53}-\rho_0
^{\frac53}-\frac{5}{3}\rho_0^{\frac23}(x+2y)\bigg\}
\end{multline*}
holds for all $x+2y\geq-\rho_0$ and all $x\geq-\rho_0$, and with
$\kappa$ as large as possible. Solving this problem numerically leads
to $t=0.7267$ and $\kappa=0.12797$.  This completes the derivation
of our first main result \eqref{eq:LT}.

Using~\eqref{eq:Legendre-V} we obtain the bound on the shift in energy
$\delta\cE(V)\geq \kappa\;\delta\cE_{\rm sc}\left({\kappa}^{-1}V\right)$.
Finally we can use the fact that $\delta\cE_{\rm
  sc}(V)-\rho_0\int_{\R^3}V(\br)\,d^3r$ is a monotone decreasing
function of $\mu$. This implies \eqref{eq:23}, which implies 
\eqref{eq:LT-dual}:
\begin{multline}  \label{eq:23}
\delta\cE_{\rm sc}\left(\frac{1}{\kappa}{V}\right)-\frac{\rho_0}\kappa\int_{\R^3}V(\br)\,d^3r\\
\geq \kappa^{-5/2}\;\left(\delta\cE_{\rm sc}(V)
-\rho_0\int_{\R^3}V(\br)\,d^3r\right) \, .
\end{multline}

\medskip
\noindent
\textbf{Extension to 2D:}
Our method can be generalized to 2D (indeed to any dimension except
1D). The result is
\begin{equation}
\delta\cT^{\rm 2D}(\delta\rho)\geq \;0.04493\; \delta\cT_{\rm sc}^{\rm 2D}(\delta\rho)
\label{eq:LT_2D}
\end{equation}
for the change in kinetic energy, and
\begin{align}
0&\geq \delta\cE^{\rm 2D}(V) - \rho_0\;\int_{\R^2} V(\br)\,d^2r \nonumber\\
 &\geq \;22.25\;\left(\delta\cE^{\rm 2D}_{\rm sc}(V)- \rho_0\;\int_{\R^2} V(\br)\,d^2r\right)
\label{eq:LT-dual_2D}
\end{align}
for the change in energy when a potential $V(\br)$ is inserted into
the system. The 2D semiclassical functions are $\cT_{\rm sc}^{\rm
  2D}(\rho)=\delta\cT_{\rm sc}^{\rm
  2D}(\rho)=(2\pi/q)\int_{\R^2}\rho(\br)^2\,d^2r$ and $\delta\cE_{\rm
  sc}^{\rm
  2D}(V)=-q/(8\pi)\int_{\R^2}\left((V(\br)-\mu)_-^2-\mu^2\right)\,d^2r$
with $\mu=4\pi\rho_0/q$.

\medskip\noindent \textbf{Peierls Instability in 1D:} In a 1D free
Fermi gas, a bound like \eqref{eq:LT-dual} cannot hold, in
general. When a potential $V(r)$ is inserted into the system at
positive density $\rho_0$, the corresponding variation of the
semiclassical energy is
$$\delta\cE^{\rm 1D}_{\rm sc}(V)=-\frac{2q}{3\pi}\int_{-\infty}^\infty\left((V(r)-\mu)^{3/2}_- - \mu^{3/2}\right)\,dr$$
with $\mu=(\pi\rho_0/q)^2$. For small $V(r)$, this gives
$$
\delta\cE^{\rm 1D}_{\rm sc}(V)-\rho_0\int_{-\infty}^\infty V(r)
\, dr\approx -\frac{q^2}{4\pi^2\rho_0}\int_{-\infty}^\infty V(r)^2\,dr.
$$ 

On the other hand, second-order perturbation theory \cite{Peierls} predicts
that the true shift $\delta\cE^{\rm 1D}(V)$, for small $V$,  is
\begin{align}
 &\delta\cE^{\rm 1D}(V)-\rho_0\int_{-\infty}^\infty V(r)\,dr\nonumber\\ 
&\qquad \qquad \approx -\frac{q^2}{2\pi}\int_{k^2\leq\mu}\,dk\int_{\ell^2\geq\mu}\,d\ell\,\frac{|\widehat{V}(k-\ell)|^2}{\ell^2-k^2}\nonumber\\
&\qquad \qquad =-\frac{q^2}{4\pi}\int_{-\infty}^{\infty}\,dk\;\frac{|\widehat{V}(k)|^2}{|k|}\log\frac{2\sqrt{\mu}+|k|}{|2\sqrt{\mu}-|k||}.
\label{eq:shift_1D}
\end{align}
The logarithm diverges at $|k|= 2\sqrt{\mu}$; hence the
second-order term can be made arbitrarily large while keeping
$\int V(r)^2\,dr$ fixed. Thus, there cannot be any
lower bound involving $\delta\cE_{\rm sc}^{\rm
  1D}(V)$.  This divergence in 1D is
well known, and is related to the Peierls
instability~\cite{Peierls}. In higher dimensions, the second-order
approximation is bounded (this follows from our
bound~\eqref{eq:LT-dual}), but it is known to have an infinite
derivative at $|k|= 2\sqrt{\mu}$, a fact that is sometimes called
the \emph{Migdal-Kohn anomaly}~\cite{Migdal}.

\medskip\noindent \textbf{Extension to positive temperature:} The
change in free energy at temperature $T=(k_B\beta)^{-1}$ and chemical
potential $\mu$ is
$$
\delta\mathcal F (V) = -\beta^{-1} \Tr\left[ \ln(1+e^{-\beta H_V}) -
\ln(1+e^{-\beta H_0}) \right]
$$
with $H_V$ and $H_0$ as before. Using the fact that $\ln(1+
e^{-\beta E}) = \beta^2 \int_E^\infty (1+ e^{\beta \lambda})^{-2}
e^{\beta\lambda} (\lambda -E)\,d\lambda$ we find
$$
\delta\mathcal F (V) = \beta \int_\R \frac{e^{\beta\lambda}}{(1+ e^{\beta \lambda})^{2}} \ \delta\mathcal E^{\mu+\lambda}(V)
 \ d\lambda
$$
with $\delta\mathcal E^{\mu+\lambda}(V)=-
\Tr\left[ (H_V-\lambda)_- - (H_0-\lambda)_- \right]$.
This formula expresses the positive temperature energy shift as a
mixture of \emph{zero temperature} energy shifts with different chemical
potentials. In fact, $\delta\mathcal E^{\mu+\lambda}(V)$ is nothing but the energy shift estimated before, with
chemical potential $\mu+\lambda$ instead of $\mu$. Thus \eqref{eq:LT-dual}
leads to
\begin{align*}
0 & \geq \delta\mathcal F(V) - \rho_T \int_{\R^3} V(\br) d^3 r \\
& \geq  -(21.85) \frac{q\beta}{15\pi^2} \int_\R d\lambda\int_{\R^3} 
 d^3r  \,  (1+ e^{\beta
\lambda})^{-2} e^{\beta\lambda} \\
&\quad \times \left\{ (V(\br)-\mu-\lambda)_-^{\frac52} -
(\mu+\lambda)_+^{\frac52} + \tfrac 52 (\mu+\lambda)_+^{\frac32} V(\br) \right\}
\end{align*}
with density
$$
\rho_T =  \frac{q}{(2\pi)^3} \int_{\R^3} (1+ e^{\beta (p^2-\mu)})^{-1}
\,d^3 p \,.
$$
Similar results hold for all $D\geq 2$.
This is a bound on the change in free energy after insertion
of a potential $V$;
the corresponding version in terms of the density change $\delta \rho$ can be
obtained via a Legendre transform, as in (\ref{eq:Legendre-rho}).

\medskip\noindent \textbf{Extension to periodic background potentials:} Our
method also works here. 
The
result depends on knowing two things: the density of states close to
the Fermi level, $\mu$, and the non-homogeneous background density
$\rho_0({\bf r})$ for this $\mu$.
With these quantities in hand, the
calculation follows along the same lines as the one given here. There are
various possible energy-band scenarios and, for lack of space, we defer the
details to a forthcoming paper.

\medskip

\noindent \textbf{Conclusion:} We show rigorously that the energy
shift of a Fermi gas, caused either by a local density perturbation
or by a local potential, is described, qualitatively, by a
semiclassical calculation.

Grants from the U.S.~NSF PHY-0965859 (E.L. and
R.F.), PHY-0845292 (R.S.) and from the ERC
MNIQS-258023 (M.L.) are  acknowledged.


\end{document}